\documentclass[12pt,a4paper]{article}

\PassOptionsToPackage{dvipsnames}{xcolor}
\usepackage{flonat-wp}
\usepackage[harvard]{flonat-bib}
\usepackage[clean]{flonat-rev}

\theoremstyle{plain}
\newtheorem{designlemma}[theorem]{Design lemma}

\title{A benchmark for strategic auditee gaming under continuous compliance monitoring}

\author{%
    Florian A.~D.~Burnat\thanks{University of Bath, Bath, England, UK. Email: \texttt{fadb20@bath.ac.uk}.}
    \and
    Brittany I.~Davidson\thanks{University of Bath, Bath, England, UK. Email: \texttt{bid23@bath.ac.uk}.}
}

\date{29 July 2026}

\begin{document}

\providecommand{\ResultDriftScheduledTau}{1.3}
\providecommand{\ResultDriftPeriodicTau}{2.0}
\providecommand{\ResultDriftOneShotTau}{5.0}
\providecommand{\ResultDriftWMin}{400}
\providecommand{\ResultDriftWMax}{550}
\providecommand{\ResultCherryGap}{0.045}
\providecommand{\ResultCherryGapSe}{0.003}
\providecommand{\ResultCherryTau}{2.7}
\providecommand{\ResultCherryBonfTau}{4.2}
\providecommand{\ResultCherryLatencyTax}{1.5}
\providecommand{\ResultAttritionCovered}{39}
\providecommand{\ResultAttritionExcluded}{534}
\providecommand{\ResultAttritionAccountSum}{573}
\providecommand{\ResultAttritionAccountIncrease}{43}
\providecommand{\ResultAttritionCoveredReduction}{90}
\providecommand{\ResultAttritionFloorTau}{2.0}
\providecommand{\ResultAttritionSuspicionTau}{12.0}
\providecommand{\ResultAttritionSuspicionCovered}{20}
\providecommand{\ResultSuspicionDriftTau}{4.0}
\providecommand{\ResultSuspicionDriftCovered}{200}
\providecommand{\ResultSuspicionDriftK}{8}
\providecommand{\ResultSuspicionCherryK}{7.0}
\providecommand{\ResultSuspicionCherryKSe}{0.36}
\providecommand{\ResultSuspicionDelayK}{3.5}
\providecommand{\ResultSuspicionDelayKSe}{0.47}
\providecommand{\ResultSuspicionAttritionDefaultK}{2.4}
\providecommand{\ResultSuspicionAttritionDefaultKSe}{0.28}
\providecommand{\ResultSuspicionAttritionActiveK}{8}
\providecommand{\ResultSuspicionHonestK}{2}
\providecommand{\ResultSuspicionOffAuditK}{2}
\providecommand{\ResultOffAuditWMin}{400}
\providecommand{\ResultOffAuditWMax}{550}
\providecommand{\ResultDefaultAttritionExcluded}{13}
\providecommand{\ResultFprPerRoundMin}{0.044}
\providecommand{\ResultFprPerRoundMax}{0.054}
\providecommand{\ResultFprFourAuditMin}{0.165}
\providecommand{\ResultFprFourAuditMax}{0.195}
\providecommand{\ResultFprBonfMin}{0.035}
\providecommand{\ResultFprBonfMax}{0.065}
\providecommand{\ResultFprPeriodicBonf}{0.040}
\providecommand{\ResultRtwoBlindTau}{1.3}
\providecommand{\ResultRtwoGuessTau}{2.1}
\providecommand{\ResultRtwoGuessOverlap}{0.283}
\providecommand{\ResultThresholdFull}{0.031}
\providecommand{\ResultThresholdSmall}{0.098}
\providecommand{\ResultThresholdFloor}{0.044}
\providecommand{\ResultExpectedFirstRandomAudit}{1.6}

\begin{titlepage}
    \maketitle

    \begin{abstract}
        \noindent Continuous post-deployment monitoring lets a regulated system shape what is reported across repeated audits. We introduce a lightweight benchmark for stress-testing such behavior in AI-governance settings. The auditor and auditee commit to complete mappings before a $T$-round rollout, following a leader--follower order; the benchmark specifies no payoffs and solves no equilibrium. Five reporting probes vary timing, reported values, evidence coverage, and audit-schedule knowledge, while five policies vary temporal coverage, sample-size conditions, and history-based escalation. An elementary lemma for the benchmark's Wald rule separates two ways a report can escape scrutiny: manipulation on unaudited rounds and weak evidence on audited rounds. Synthetic experiments show that a sample-size floor and escalation address these failures differently, and that OffAuditDrift's zero audited gap disappears when a randomized schedule is kept private. We report covered- and excluded-surface exposure accounts, not social welfare. Parameters combine institutional anchors, scale-informed choices, and mechanism-activating stress-test points drawn from the DSA context; they are not fitted to platform trajectories. The artifact reproduces the full grid, honest-noise calibration, and private-schedule diagnostic in under a minute on a laptop.

        \bigskip
        \noindent\textbf{Keywords:} algorithmic auditing, AI accountability, continuous compliance, Digital Services Act, strategic reporting, benchmark.
    \end{abstract}

    \setcounter{page}{0}
    \thispagestyle{empty}
\end{titlepage}

\clearpage
\doublespacing

\section{Introduction}
\label{sec:intro}

The European Union's AI Act mandates \emph{post-market monitoring} for high-risk AI systems (Article~72): providers must continuously document and analyze their systems' real-world behavior throughout the system's lifetime, not merely at the point of release. The Digital Services Act (DSA) imposes related governance obligations on large online platforms (VLOPs/VLOSEs), including transparency reports on content moderation (Article~15), statements of reasons for moderation actions (Article~24), six-month publication cycles for VLOPs (Article~42), systemic-risk assessment (Article~34) and mitigation (Article~35), and annual independent audits (Article~37). These regimes share a common structural feature: \emph{the audit relationship is longitudinal}. A regulated firm reports outcome metrics across many rounds, an auditor decides which rounds to scrutinize, and the firm \new{can produce time- and history-dependent outputs from its precommitted mapping}\removed{can adapt over time}.

This longitudinal structure creates a class of strategic auditee behavior that prior work on AI auditing has largely sidestepped. The strategic classification literature \parencite{Hardt2016-vu, Dong2018-ht}, performative prediction \parencite{Perdomo2020-gp}, and audit games \parencite{Blocki2015-du} all model essentially \emph{one-shot} interactions: a single classifier deployed against strategic agents, or a single audit allocation across one population. Recent work on AI audit ecosystems and access requirements \parencite{Raji2022-ks, Casper2024-em} has clarified what auditors need but assumed an honest-but-curious auditee. The most directly comparable empirical study---a deployed multi-party computation protocol for fairness monitoring \parencite{He2026-it}---explicitly assumes honest-but-curious parties. None of this work asks how a \new{strategic}\removed{n adaptive} auditee should be expected to behave when audited continuously, with discretion over the timing, definition, and population coverage of its reports.

Empirical work on continuous compliance reveals that this gap is not merely theoretical. An audit of the DSA Transparency Database documents inconsistencies in moderation reporting across platforms \parencite{Trujillo2025-yz}, and recent work documents platform-side restrictions on API access that create ``audit blind-spots'' for independent verification across X/Twitter, Reddit, TikTok, and Meta \parencite{Burnat2026-xq}. Continuous-auditing research in accounting has long acknowledged strategic auditee behavior \parencite{Heald2018-hj} but has not been translated into shared, executable evaluation primitives for the AI-auditing community.

\new{The mechanisms studied here are not uniquely AI-specific: schedule exploitation, evidence-base manipulation, and metric-definition discretion arise in continuous compliance generally. AI specificity enters through the monitoring application and DSA/AI Act institutional context. Accordingly, our claims concern a synthetic stress-test environment, not a validated behavioral model of regulated platforms.}

\paragraph{Contributions.}
\begin{itemize}
  \item \textbf{(C1) Process specification.} We formulate continuous compliance as a $T$-round stochastic audit process with a leader--follower commitment order over complete mappings. The timing is inspired by Stackelberg commitment, but the benchmark specifies no payoffs and solves no Stackelberg equilibrium. Outputs can depend on time and history, but neither player learns or recomputes a best response during a rollout.
  \item \textbf{(C2) Two exposure accounts.} The non-additive pair $(W,C)$ separates population-weighted misrepresentation inside the verified report on unaudited rounds from misrepresentation affecting the population excluded from that report. The quantities diagnose a shift between accountability surfaces; they are not welfare estimates and are not meaningfully aggregated.
  \item \textbf{(C3) A Wald-regime design lemma.} Design Lemma~\ref{prop:no-static} identifies a coverage--granularity trade-off for a precisely scoped static temporal-coverage class and motivates one intervention along each failure axis.
  \item \textbf{(C4) A diagnostic library and reproducible artifact.} Five auditee probes span timing, reported value and definition, evidence coverage and precision, and schedule knowledge. Three temporal-coverage baselines are followed by a precision-conditioned intervention and a history-conditioned intervention. The library supports conditional stress tests; it is neither an exhaustive strategy space nor an equilibrium solution.
  \item \textbf{(C5) Synthetic stress-test experiments.} We distinguish an institutionally anchored horizon, scale-informed illustrative choices, and mechanism-activating operating points motivated by a published DSA Transparency Database audit \parencite{Trujillo2025-yz}. We also report honest-noise calibration and a private-draw schedule diagnostic. The benchmark does not ingest or fit platform-level trajectories.
\end{itemize}

\section{Related work}
\label{sec:related}

\paragraph{Strategic classification and performative prediction.}
Strategic classification studies agents who modify features to game a classifier, including revealed-preference settings \parencite{Hardt2016-vu, Dong2018-ht}. Performative prediction instead captures population-level shifts induced by deployment \parencite{Perdomo2020-gp}. Both remain essentially one-shot: a single classifier is deployed against strategic agents, with no audit cadence chosen by an auditor. We study the dual setting, where the strategic agent is the regulated system itself and the auditor commits to a temporal policy.

\paragraph{Audit games.}
Multi-resource audit games formalize the allocation of audit effort across multiple targets under a budget constraint \parencite{Blocki2015-du}. We borrow the bilevel-with-commitment template but specialize to a longitudinal compliance setting in which a single regulator--firm pair interacts over $T$ reporting cycles, and where the strategic move is the auditee's choice of \emph{when} and \emph{how much} to drift, not the allocation of defender resources.

\paragraph{Privacy-preserving fairness monitoring.}
The most directly comparable applied work provides an MPC-based fairness monitoring protocol for algorithmic hiring, validated in a real industrial deployment \parencite{He2026-it}. That framework assumes honest-but-curious parties and uses a single-round audit interface. Recent game-theoretic treatments of privacy-preserving algorithmic accountability \parencite[e.g.,][]{Yang2025-lv} also remain single-round. Our benchmark is the temporal/strategic complement: it evaluates policy-aware reporting probes over longitudinal trajectories and compares detection latency with the two exposure accounts.

\paragraph{AI audit institutions.}
Research on third-party audit ecosystems, black-box access, and platform API restrictions clarifies what regulators need from access and information rights \parencite{Raji2022-ks, Casper2024-em, Burnat2026-xq}, but stops short of executable longitudinal stress tests for policy-aware reporting. Complementary political-economy work endogenizes vendor auditability, mitigation, and deployer monitoring under vendor--deployer lock-in and evidence-dependent enforcement \parencite{Burnat2026-jd}; the present benchmark instead leaves incentives unspecified and stress-tests reporting mappings longitudinally. Empirical work on the DSA Transparency Database documents inconsistencies in self-reported moderation actions but does not identify strategic intent \parencite{Trujillo2025-yz}. Our benchmark supplies a synthetic environment in which analysts can test specified reporting probes against specified audit policies under common metrics.

\paragraph{NeurIPS Datasets \& Benchmarks precedents.}
HELM \parencite{Bommasani2023-lz} exemplifies the format for benchmark contributions in this venue: a documented evaluation environment, a taxonomy of test cases, baseline implementations, and reproducibility scaffolding. Our benchmark adopts this format in a regulatory rather than capability-evaluation domain and pairs it with a minimum-bar reproducibility design: single-command installation, deterministic seeds, and JSON outputs.

\paragraph{Continuous auditing in accounting.}
A long-running literature in accounting and public administration has formalized continuous auditing as a sister problem \parencite{Shore2015-gc, Heald2018-hj, Parker2021-tg}. This literature acknowledges strategic auditee behavior but has not produced shared evaluation primitives that the AI fairness community has adopted. Our benchmark is in part a translation effort: porting the continuous-auditing intuition into a runnable, ML-friendly environment.

\section{The benchmark}
\label{sec:benchmark}

\subsection{Audit process and commitment}
\label{sec:setup}

\new{We model continuous compliance as a $T$-round stochastic audit process with a leader--follower commitment order between an \emph{auditor} and an \emph{auditee}. At round $t \in \{0,1,\dots,T-1\}$:}

\begin{enumerate}
  \item The latent full-population quantity $m_t^\star \in [0,1]$ evolves according to clipped baseline drift, $m_t^\star = \mathrm{clip}(m_{t-1}^\star + \eta_t,0,1)$ with $\eta_t \sim \mathcal{N}(0,\sigma^2)$ and $\sigma=0.02$ in the experiments.
  \item The auditee's committed reporting mapping outputs a reported quantity $\hat m_t \in [0,1]$ and an included population or effective evidence-base size $n_t \in \{n_{\min},n_{\max}\}$. Its output can depend on time, the disclosed audit information, and observed history.
  \item The auditor's committed policy mapping outputs an audit decision. A static mapping conditions only on time; a history-conditioned mapping can also condition on prior observed reports and audit outcomes.
\end{enumerate}

\new{\paragraph{Commitment and equilibrium scope.}The auditor commits a complete policy mapping first, and the auditee then commits a complete reporting mapping given the information disclosed at that point. The mappings remain fixed during rollout even when their realized outputs depend on history. Thus the commitment stage admits a one-shot leader--follower representation over complete contingent plans, while the $T$-round rollout remains necessary for stochastic drift, detection latency, accumulated exposure, and history-conditioned audit actions. This timing is inspired by Stackelberg commitment, but the benchmark does not specify payoffs or compute utilities, best responses, or a Stackelberg equilibrium: each supplied auditee mapping is a conditional adversarial probe.}

\new{\paragraph{Schedule formation and disclosure.}We separate two axes that the earlier R1/R2/R3 shorthand conflated. \emph{Formation} asks whether audit decisions are fixed deterministically, drawn from a committed distribution, or produced by a history-conditioned mapping. \emph{Disclosure} asks whether the policy and any realized fixed schedule are known to the auditee. The headline One-shot, Periodic, Scheduled-random, and Min-sample-floor cells disclose their realized schedules, so Scheduled-random changes coverage geometry but creates no strategic surprise. Suspicion-escalation discloses its mapping, but its realized audit set is history-dependent. We retain \emph{R1} for a disclosed realized schedule and \emph{R2} for a public randomized policy with a private realized draw; \emph{R3} denotes private or richer history-conditioned scheduling and remains outside the evaluated grid.}

The auditor's decision rule on audited rounds is a noise-aware Wald-style threshold (with regulatory tolerance $\epsilon \geq 0$ and $z = 1.96$ for a 95\% Wald CI half-width): a gap is flagged when
\begin{equation}
\label{eq:detection-rule}
|\hat m_t - m_t| > \epsilon + z \cdot \sqrt{\hat m_t (1 - \hat m_t) / n_t}.
\end{equation}
\new{\paragraph{Reference and estimand semantics.}In the current simulator, the auditor's verified reference $m_t$ is set equal to the latent full-population quantity $m_t^\star$. The auditee reports $\hat m_t$ with included-population or evidence-base size $n_t$. Except for the honest-noise control, strategies generate $\hat m_t$ directly rather than drawing an individual-level sample of size $n_t$. The rule is therefore a deliberately simplified acceptance band around the report, not a full two-sample test; a smaller $n_t$ both widens the band and narrows the population to which the report speaks.}

\new{\paragraph{Multiplicity correction.}The uncorrected rule uses $z=1.96$. Its family-wise variant substitutes $z_{\mathrm{Bonf}}=\Phi^{-1}(1-\alpha/(2K))$, where $\Phi$ is the standard normal cumulative distribution function, $\alpha=0.05$, and $K$ is the number of audited rounds. For fixed schedules, $K$ is the committed audit-set size; for Suspicion-escalation, it is the realized per-trajectory size (Appendix~\ref{app:bonferroni-K}). We report both variants because honest-noise calibration shows that the correction controls family-wise false positives while increasing detection latency for stochastic-gap probes.}

\new{\paragraph{Preview of the design lemma.}Two failure modes organize everything downstream. Under a fixed temporal schedule that uses rule~\eqref{eq:detection-rule} as its sole acceptance criterion, a moderate misreport can escape through \emph{coverage failure} (Case~1), by appearing only on unaudited rounds, or through \emph{granularity failure} (Case~2), by appearing with a sufficiently small $n_t$. Design Lemma~\ref{prop:no-static} (\S\ref{sec:theory}) scopes this claim to the Wald-rule regime used here.}

\subsection{Auditee gaming strategies}

\new{\paragraph{Diagnostic-suite selection.}The auditee library is a small diagnostic suite spanning the simulator's reporting degrees of freedom: timing (Delay), reported value and metric construction (Drift and Cherry-pick), population coverage and evidentiary precision (Attrition), and knowledge of the audit schedule (OffAuditDrift). Each primitive isolates one mechanism so that strategy--policy cells remain interpretable. The library is neither exhaustive nor an estimate of how often regulated firms use these behaviors.}

\begin{itemize}
  \item \textbf{Delay} ($k = 2$): the auditee reports the metric value from $k$ rounds prior, lagging unfavorable shifts. Models reporting frictions and the strategic use of submission cycles to push bad news beyond the regulator's attention window. In our (symmetric-drift) experiments the gap can take either sign; in adverse-shift regimes (e.g., monotonically degrading systems) Delay produces consistently positive gaps.
  \item \textbf{Drift} ($\delta = 0.05$): the auditee adds a fixed bias to the truth, reporting $\hat m_t = m_t + \delta$. Models the strategic choice of a single fixed operationalization of an ambiguous regulatory metric.
  \item \textbf{Cherry-pick} ($J = 5$ candidates, $\sigma_{\text{pick}} = 0.04$): a generic definition-shopping probe. Each round, the auditee draws $J$ candidate metric estimates from $\mathcal{N}(m_t,\sigma_{\text{pick}}^2)$ and reports the maximum. This stylizes settings in which defensible denominators, subgroup definitions, or aggregation rules yield different reportable values; regulatory discretion is motivating context, not evidence that DSA Article~15 specifies this exact problem. Cherry-pick is Drift's stochastic-gap counterpart and makes detection latency sensitive to the Bonferroni choice (Table~\ref{tab:headline}).
  \item \textbf{Attrition} ($\tau = 0.40$, $\delta_{\text{cover}} = 0.05$): when $m_t < \tau$ the auditee restricts to a smaller verified sub-sample, dropping $n_t$ from $n_{\max} = 1000$ to $n_{\min} = 100$, and additionally drifts the report by $\delta_{\text{cover}}$. The reduced sample widens the auditor's tolerance band per~\eqref{eq:detection-rule}, allowing a small drift to escape.
  \item \textbf{OffAuditDrift} ($\delta = 0.05$): a schedule-aware variant of Drift. When the realized audit set $\mathcal{A}_\pi$ is disclosed at commitment, the auditee reports honestly ($\hat m_t=m_t$) on those rounds and drifts by $\delta$ elsewhere. This instantiates Design Lemma~\ref{prop:no-static}, Case~1, for fixed schedules. Against Suspicion-escalation it conditions on the disclosed base cadence; because those audits observe zero gap, escalation never activates. The R2 diagnostic withholds the realized randomized schedule.
\end{itemize}

A single common interface (\verb|choose_report(t, m_t, history) -> (m_hat_t, n_t)|) lets users compose, parameterize, or extend these. We provide an \texttt{HonestAuditee} baseline that returns $(m_t, n_{\max})$ unconditionally as the no-strategic counterfactual.

\subsection{Auditor policies}

\new{The auditor library follows the same diagnostic logic: three temporal-coverage baselines, followed by one precision-conditioned intervention and one history-conditioned intervention. The first two baselines have broad regulatory analogies, while the latter interventions are benchmark proposals rather than requirements currently prescribed by the DSA or AI Act.}

\begin{itemize}
  \item \textbf{One-shot} ($t^\star = 5$): audit at a single committed round $t^\star$. This stylizes a point-in-time inspection such as an independent audit.
  \item \textbf{Periodic} (period $k = 3$, phase $k - 1$): audit at $t \in \{2,5,8,11\}$ over $T=12$. This stylizes recurring reporting and monitoring cycles under DSA Articles~15/42 and AI Act Article~72; it is not a claim that those instruments prescribe this exact benchmark cadence.
  \item \textbf{Scheduled-random} ($K=4$ rounds, seed $42$): audit at $K$ pseudo-randomly selected rounds. In the headline R1 cells the realized schedule is disclosed, making this a control that changes coverage geometry without creating secrecy. The R2 diagnostic keeps the same public distribution but withholds the realized draw.
  \item \textbf{Min-sample floor} ($n_{\text{floor}}=500$, period $k=3$, phase $k-1$): a proposed benchmark intervention that audits on the Periodic cadence and flags an audited report whenever $n_t<n_{\text{floor}}$, in addition to the Wald rule. It conditions on evidentiary precision and is not presented as an existing regulatory requirement.
  \item \textbf{Suspicion-escalation} (base period $4$, suspicion threshold $0.04$): a proposed history-conditioned intervention. It starts from base-cadence audits at $t\in\{4,8\}$ and, after an observed gap above the threshold, audits every subsequent round. The committed mapping is fixed, but its realized audit set depends on history.
\end{itemize}

The first three policies fall inside the static temporal-coverage class of Design Lemma~\ref{prop:no-static}. Min-sample floor leaves that class by conditioning acceptance on $n_t$, and Suspicion-escalation leaves it by conditioning future audits on observed history. Richer Bayesian, sequential, or learned policies remain follow-on work.

\subsection{Evaluation metrics}
\label{sec:metrics}

We report four primary quantities on each auditee--auditor pair, averaged over $S$ random seeds:

\begin{itemize}
  \item \textbf{Gaming gap} $G(\pi, \sigma) = \overline{(\hat m_t - m_t)}_{t \in \mathcal{A}}$, the mean reported-vs-true difference over audited rounds $\mathcal{A}$. Positive values indicate over-reporting (favorable to the auditee).
  \item \textbf{Time-to-detection} $\tau_d(\pi,\sigma) \in \{0,1,\dots,T\}$: the first audited round at which $|\hat m_t-m_t|$ exceeds the decision threshold; $\tau_d=T$ encodes no detection within the horizon. We report the uncorrected and Bonferroni variants defined above. Detection does not change the auditee's committed mapping.
  \item \textbf{Covered-surface exposure} $W(\pi,\sigma)=\sum_{t\notin\mathcal{A}} n_t|\hat m_t-m_t|$: population-weighted misrepresentation inside the report's included surface, accumulated over unaudited rounds.
  \item \textbf{Excluded-surface exposure} $C(\pi,\sigma)=\sum_t (N-n_t)|\hat m_t-m_t|$: population-weighted misrepresentation affecting the population excluded from the report, with $N=n_{\max}$ in the headline tables.
\end{itemize}

\new{The implementation and JSON use \texttt{covered\_surface\_exposure} and \texttt{excluded\_surface\_exposure} as the primary names; \texttt{welfare\_loss} and \texttt{coverage\_loss} remain documented legacy aliases for backward compatibility. Neither quantity is a statistically estimated welfare loss. They describe disjoint accountability surfaces, so $W+C$ is not a sufficient harm statistic; the informative comparison is how Attrition shifts weighted exposure from the covered to the excluded surface.}

\section{A design lemma for the Wald regime}
\label{sec:theory}

\new{The following elementary lemma delimits the static-policy and Wald-rule regime studied by the benchmark. Its role is organizational: it separates two failure axes and motivates one baseline intervention along each axis, rather than claiming a general impossibility result.}

We restrict attention to \emph{static temporal-coverage policies}: policies $\pi$ that specify an audit set $\mathcal{A}_\pi \subseteq \{0,\dots,T-1\}$ at commitment, condition only on the round index, and use rule~\eqref{eq:detection-rule} as their sole acceptance criterion. This is the class containing the first three benchmark policies. DSA and AI Act obligations motivate longitudinal monitoring, but they do not prescribe this exact policy class or acceptance rule. Min-sample floor and Suspicion-escalation are proposed diagnostic interventions outside the class.

\begin{designlemma}[Coverage--granularity trade-off in the cover regime]
\label{prop:no-static}
Fix horizon $T \geq 2$, regulatory tolerance $\epsilon \geq 0$, and the noise-aware decision rule~\eqref{eq:detection-rule} with sample-size bounds $n_{\min} < n_{\max}$. Define the \emph{cover regime} at $\hat m \in (0, 1)$ as the half-open interval
\begin{equation*}
\Delta(\hat m, n_{\min}, n_{\max})
\;=\;
\bigl(
\epsilon + z\sqrt{\hat m(1-\hat m)/n_{\max}},\;
\epsilon + z\sqrt{\hat m(1-\hat m)/n_{\min}}
\bigr],
\end{equation*}
i.e., drift magnitudes large enough to exceed the full-sample threshold but small enough to fit within the small-sample threshold (worked example below). Define $h(p, n) = \epsilon + z\sqrt{p(1-p)/n}$. For any static temporal-coverage policy $\pi$ and any $\delta$ such that $\delta \in \Delta(\hat m_t, n_{\min}, n_{\max})$ uniformly in $t$ (i.e., $\max_t h(\hat m_t, n_{\max}) < \delta \leq \min_t h(\hat m_t, n_{\min})$ across the realized trajectory), an evading auditee strategy $\sigma$ exists in either of two structurally distinct configurations:
\begin{enumerate}
  \item \textbf{Coverage failure.} If $\mathcal{A}_\pi \subsetneq \{0, \dots, T-1\}$, strategy $\sigma_1$ drifts by $\delta$ at any unaudited round $t^\star \notin \mathcal{A}_\pi$. The drift is never observed, so $\tau_d(\pi, \sigma_1) = T$ and covered-surface exposure $W(\pi, \sigma_1) \geq \delta \cdot n_{\max} > 0$.
  \item \textbf{Granularity failure.} If $\mathcal{A}_\pi = \{0, \dots, T-1\}$, an attrition-style mapping $\sigma_2$ reports $n_t = n_{\min}$ and drifts by $\delta$ on every relevant round. Unlike the implemented threshold-activated Attrition probe, this constructive mapping does not require $m_t<\tau$. The threshold widens beyond $\delta$, so the gap is never flagged: $\tau_d(\pi, \sigma_2) = T$ and $G(\pi, \sigma_2) = \delta > 0$.
\end{enumerate}
The trade-off binds this \emph{static temporal-coverage} class: auditing more often reduces Case~1 but does not impose an independent evidence-size constraint for Case~2. A minimum-sample-size floor lies outside the class and can defeat Case~2 directly.
\end{designlemma}

\begin{proof}
\emph{Case 1.} Suppose $\mathcal{A}_\pi \subsetneq \{0, \dots, T-1\}$ and pick any $t^\star \notin \mathcal{A}_\pi$. Define $\sigma_1$ to drift by $\delta$ at $t^\star$ and report honestly otherwise: $\hat m_{t^\star} = m_{t^\star} + \delta$ and $\hat m_t = m_t$ for $t \neq t^\star$, with $n_t = n_{\max}$ throughout. Detection requires $|\hat m_t - m_t|$ to exceed the threshold of~\eqref{eq:detection-rule} on some round in $\mathcal{A}_\pi$; but on every $t \in \mathcal{A}_\pi$ we have $|\hat m_t - m_t| = 0$, so detection cannot fire and $\tau_d(\pi, \sigma_1) = T$. The exposure bound follows from $W(\pi, \sigma_1) \geq |\hat m_{t^\star} - m_{t^\star}| \cdot n_{\max} = \delta \cdot n_{\max} > 0$.

\emph{Case 2.} Suppose $\mathcal{A}_\pi = \{0, \dots, T-1\}$ and consider $\sigma_2$ with $n_t = n_{\min}$ and $\hat m_t = m_t + \delta$ on every round. The threshold of~\eqref{eq:detection-rule} at round $t$ is $h(\hat m_t, n_{\min})$. By the uniform hypothesis $\delta \leq \min_t h(\hat m_t, n_{\min})$, so $|\hat m_t - m_t| = \delta \leq h(\hat m_t, n_{\min})$ at every round; the rule does not fire and $\tau_d(\pi, \sigma_2) = T$. The audited-round gap satisfies $G(\pi, \sigma_2) = \mathbb{E}_{t \in \mathcal{A}_\pi}[\hat m_t - m_t] = \delta > 0$ by construction. (The uniform condition is automatic when the trajectory $\hat m_t$ stays bounded away from $0$ and $1$, as in the empirical anchor with $\hat m \in [0.30, 0.55]$; near the boundary, $p(1-p)$ shrinks and the regime can close.)

The proof is constructive and immediate from the Wald threshold. The benchmark contribution is the executable diagnostic organization around these two cases.
\end{proof}

\paragraph{Reading the lemma.} The cover regime is non-empty whenever $n_{\max}/n_{\min}$ is bounded away from~$1$; at the benchmark parameters ($n_{\min}=100$, $n_{\max}=1000$, $\hat m=0.5$) it is $(\ResultThresholdFull,\ResultThresholdSmall]$ and contains $\delta=0.05$. Outside that interval the lemma is silent. The two intervention baselines demonstrate class changes along the evidence-size and audit-history axes; they are not claims of optimal policy. Figure~\ref{fig:sensitivity} maps the regime in $(n_{\min},\delta)$ space for Periodic.

\section{Empirical anchor: content moderation under the EU DSA}
\label{sec:case-study}

\paragraph{Parameterization.} DSA content-moderation obligations span Articles~15 (transparency reports), 24(5) (statements of reasons), 34--35 (systemic-risk assessment / mitigation), 37 (independent audits), and 42 (six-month cycles for VLOPs). Table~\ref{tab:calibration} separates an institutional timing anchor from scale-informed context and mechanism-activating stress-test choices. The benchmark does not ingest DSA-TDB records or estimate these parameters from platform trajectories (see \S\ref{sec:discussion}).

\begin{table}[h]
\centering
\footnotesize
\caption{Benchmark parameterization. ``Institutional'' denotes a direct timing analogy; ``stress test'' denotes an illustrative operating point chosen to activate a benchmark mechanism; ``statistical'' denotes an experimental convention. DSA-TDB evidence supplies context, not a per-platform fit.}
\label{tab:calibration}
\begin{tabular}{p{0.16\linewidth} p{0.11\linewidth} p{0.15\linewidth} p{0.46\linewidth}}
\toprule
Parameter & Default & Role & Rationale \\
\midrule
$T$ horizon & $12$ & Institutional & Six-year synthetic horizon at the DSA Article~42 six-month reporting cadence. \\
$m_0$ baseline & $0.5$ / $0.30$ & Stress test & $0.50$ centers the bounded state; $0.30$ keeps the implemented Attrition condition $m_t<0.40$ active. Neither is inferred as a parity estimate from DSA records. \\
$\delta$ drift & $0.05$ & Stress test & Mechanism-activating point inside the cover interval. DSA reporting variability motivates definition discretion but does not identify a five-point strategic bias \parencite{Trujillo2025-yz}. \\
$n_{\max}, n_{\min}$ & $1000, 100$ & Stress test & Tenfold contrast chosen to widen the Wald band. Inter-platform Statements-of-Reasons volume heterogeneity motivates scale variation but does not establish tenfold within-report discretion \parencite{Trujillo2025-yz}. \\
$\epsilon, z$ & $0, 1.96$ & Statistical & Benchmark baseline using a two-sided 95\% Wald band; not a claim about current regulatory practice. \\
$S$ seeds & $30$ & Statistical & Deterministic seed set; tables report mean $\pm$ SE. \\
\bottomrule
\end{tabular}
\end{table}

\paragraph{Results.}
Table~\ref{tab:headline} reports the strategy-by-policy results for both configs.

\begin{table}[t]
\centering
\caption{Headline results, 30-seed averages (seeds $0$--$29$). $\tau_d^{\text{uncorr}}$, $\tau_d^{\text{Bonf}}$: time-to-detection under per-round 95\% Wald-CI ($z=1.96$) vs.\ Bonferroni correction ($\alpha=0.05$ over $K$ audited rounds; see Appendix~\ref{app:bonferroni-K}); $\tau_d = 12.0$ means ``never detected within horizon $T=12$''. $W, C$: two disjoint accountability surfaces (\S\ref{sec:case-study}), not summands of a single total. The Honest row collapses all five policies into one (identical zero-gap, never-detect, zero-exposure row).}
\label{tab:headline}
\small
\begin{tabular}{llrrrrr}
\toprule
\textbf{Strategy} & \textbf{Policy} & \textbf{Gap} & \textbf{$\tau_d^{\text{uncorr}}$} & \textbf{$\tau_d^{\text{Bonf}}$} & \textbf{$W$} & \textbf{$C$} \\
\midrule
\multicolumn{7}{l}{\emph{Default config ($m_0=0.5$, attrition rarely activates)}} \\
Honest & \emph{any of the five} & $\phantom{+}0.000 \pm 0.000$ & $12.0$ & $12.0$ & $0$ & $0$ \\
Drift & One-shot & $+0.050 \pm 0.000$ & $5.0$ & $5.0$ & $550 \pm 0$ & $0$ \\
Drift & Periodic & $+0.050 \pm 0.000$ & $2.0$ & $2.0$ & $400 \pm 0$ & $0$ \\
Drift & Scheduled-random & $+0.050 \pm 0.000$ & $1.3$ & $1.3$ & $400 \pm 0$ & $0$ \\
Drift & Min-sample floor & $+0.050 \pm 0.000$ & $2.0$ & $2.0$ & $400 \pm 0$ & $0$ \\
Drift & Suspicion-escalation & $+0.050 \pm 0.000$ & $\mathbf{4.0}$ & $\mathbf{4.0}$ & $200 \pm 0$ & $0$ \\
OffAuditDrift & One-shot & $\phantom{+}0.000 \pm 0.000$ & $12.0$ & $12.0$ & $\mathbf{550}$ & $0$ \\
OffAuditDrift & Periodic & $\phantom{+}0.000 \pm 0.000$ & $12.0$ & $12.0$ & $\mathbf{400}$ & $0$ \\
OffAuditDrift & Scheduled-random & $\phantom{+}0.000 \pm 0.000$ & $12.0$ & $12.0$ & $\mathbf{400}$ & $0$ \\
OffAuditDrift & Min-sample floor & $\phantom{+}0.000 \pm 0.000$ & $12.0$ & $12.0$ & $\mathbf{400}$ & $0$ \\
OffAuditDrift & Suspicion-escalation & $\phantom{+}0.000 \pm 0.000$ & $12.0$ & $12.0$ & $\mathbf{500}$ & $0$ \\
Cherry-pick & Periodic & $+0.045 \pm 0.003$ & $2.7$ & $\mathbf{4.2}$ & $375 \pm 13$ & $0$ \\
Cherry-pick & Scheduled-random & $+0.045 \pm 0.002$ & $2.3$ & $2.8$ & $379 \pm 14$ & $0$ \\
Cherry-pick & Min-sample floor & $+0.045 \pm 0.003$ & $2.7$ & $4.2$ & $375 \pm 13$ & $0$ \\
Delay & Periodic & $-0.002 \pm 0.002$ & $6.4$ & $\mathbf{9.0}$ & $163 \pm 12$ & $0$ \\
\midrule
\multicolumn{7}{l}{\emph{Attrition config ($m_0 = 0.30$, attrition fires every round)}} \\
Attrition & One-shot & $+0.050 \pm 0.000$ & $12.0$ & $12.0$ & $54 \pm 0$ & $\mathbf{534} \pm 4$ \\
Attrition & Periodic & $+0.050 \pm 0.000$ & $12.0$ & $12.0$ & $39 \pm 0$ & $534 \pm 4$ \\
Attrition & Scheduled-random & $+0.049 \pm 0.001$ & $12.0$ & $12.0$ & $40 \pm 0$ & $534 \pm 4$ \\
Attrition & Min-sample floor & $+0.050 \pm 0.000$ & $\mathbf{2.0}$ & $\mathbf{2.0}$ & $39 \pm 0$ & $534 \pm 4$ \\
Attrition & Suspicion-escalation & $+0.049 \pm 0.001$ & $12.0$ & $12.0$ & $20 \pm 0$ & $534 \pm 4$ \\
\bottomrule

\end{tabular}
\end{table}

Five patterns dominate the table.

\textbf{(1) Drift is detected, but unaudited exposure remains.} Against Drift, detection latency follows audit cadence: Scheduled-random detects in $\ResultDriftScheduledTau$ rounds on average, Periodic in $\ResultDriftPeriodicTau$, and One-shot in $\ResultDriftOneShotTau$. For $K=4$ draws from $T=12$, the analytical expected first audited round is $(T+1)/(K+1)-1=\ResultExpectedFirstRandomAudit$, close to the finite-seed Scheduled-random mean. Above-envelope drift is detected, but without remediation the gap persists on unaudited rounds, producing covered-surface exposure $W=\ResultDriftWMin$--$\ResultDriftWMax$ person-rounds. OffAuditDrift below is the direct Case~1 witness.

\textbf{(2) Bonferroni control trades false positives for latency.} Cherry-pick's Periodic mean gap is $+\ResultCherryGap$ with seed-level SE $\ResultCherryGapSe$. The uncorrected rule detects in $\ResultCherryTau$ rounds, while Bonferroni correction across $K=4$ audits increases the mean to $\ResultCherryBonfTau$, a $\ResultCherryLatencyTax$-round latency difference. Drift's fixed gap remains above both thresholds. The honest-noise control below shows why the correction matters statistically.

\begin{figure}[!htb]
\centering
\includegraphics[width=0.58\linewidth]{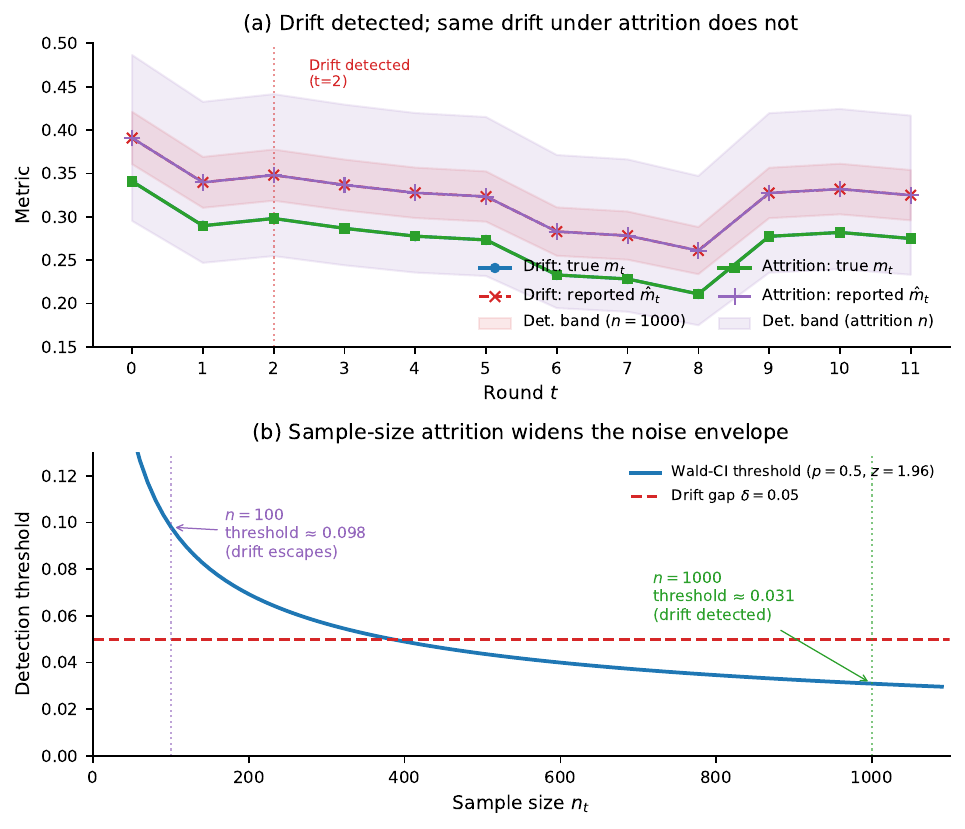}
\caption{\textbf{Why attrition escapes detection.}
(a) Per-round trajectories under periodic audit ($k=3$): Drift (red) and Attrition (purple) both produce a $+0.05$ gap, but Drift's report at $n_t = 1000$ sits outside the narrow noise band (flagged on first audit), while Attrition's $n_t = 100$ widens the band so the same gap fits inside (never flagged).
(b) Wald-CI 95\% detection threshold as a function of $n_t$ at $\hat m = 0.5$ (red dashed line: $\delta = 0.05$): threshold $\approx \ResultThresholdFull$ at $n_t = 1000$ (drift detected); $\approx \ResultThresholdSmall$ at $n_t = 100$ (drift swallowed).}
\label{fig:headline}
\end{figure}

\textbf{(3) Attrition shifts the measured surface.} Against Attrition, all three static policies fail to detect under either rule despite a $+0.05$ gap because $n_t$ falls from $1000$ to $100$, widening the threshold from about $\ResultThresholdFull$ to $\ResultThresholdSmall$ (Figure~\ref{fig:headline}). Under Periodic, Drift yields $W=\ResultDriftWMin$ and $C=0$, whereas Attrition yields $W=\ResultAttritionCovered$ and $C=\ResultAttritionExcluded$. This is not evidence that total social harm rises or falls: it shows that the benchmark's weighted exposure shifts from the report's covered surface to its excluded surface, as in Design Lemma~\ref{prop:no-static}, Case~2.

\textbf{(4) The two interventions address different axes.} Min-sample floor conditions on $n_t$ and detects Attrition on its first Periodic audit, but it would also flag a legitimately small population below the same floor. Suspicion-escalation conditions on observed gap history. Against Drift it detects in $\ResultSuspicionDriftTau$ rounds and reduces covered-surface exposure to $\ResultSuspicionDriftCovered$. Against Attrition in the attrition configuration, the first base audit observes a raw gap above $0.04$ and escalation expands the realized audit set to $K=\ResultSuspicionAttritionActiveK$, reducing $W$ to $\ResultAttritionSuspicionCovered$; formal detection still fails because the same gap remains inside the widened Wald band at $n_t=100$. Neither intervention is a universal solution or a claimed current regulatory requirement.

\textbf{(5) OffAuditDrift exposes the public-schedule condition.} When the realized schedule is disclosed, OffAuditDrift makes the audited-round gap zero by construction, so no headline policy detects it and covered-surface exposure ranges from $\ResultOffAuditWMin$ to $\ResultOffAuditWMax$. For Suspicion-escalation, zero gaps on the disclosed base cadence prevent escalation. This is a conditional Case~1 witness, not a claim about private or secret schedules; the R2 diagnostic below changes that information condition.

\paragraph{Reproducibility.} Pure-Python package (Python $\geq 3.11$); a single-seed cell runs in $<1$\,s, the full 30-seed sweep ($60$ cells) in $<1$\,min. Reproduction commands, three YAML configurations, precomputed outputs, and the 23-test pytest suite are in Appendix~\ref{app:implementation}. Regression tests pin the corrected Suspicion-escalation cells, the three R2 rows, and selected headline values; the MIT-licensed review mirror is at \url{https://anonymous.4open.science/r/neurips-2026-audit-gaming-artifact-8FDD/}.

\subsection{Validation checks}
\label{sec:fpr-r2}

\paragraph{Honest-noise calibration.}
\new{Under \texttt{HonestNoisyAuditee}, the uncorrected per-round false-positive rate is $\ResultFprPerRoundMin$--$\ResultFprPerRoundMax$ at nominal $\alpha=0.05$; Bonferroni holds the family-wise rate at or below $\ResultFprBonfMax$ while increasing latency for Cherry-pick (Appendix~\ref{app:fpr}).}

\paragraph{R2 private-draw diagnostic.}
\new{The headline OffAuditDrift result gives the reporting rule the realized Scheduled-random audit set. R2 removes that disclosure while preserving the same horizon and 30 environment seeds. Because the benchmark does not specify auditee payoffs, we evaluate two simple conditional substitute probes rather than label either a best response: unconditional drift, and honest reporting on an independently guessed schedule.}

\begin{table}[ht]
\centering
\footnotesize
\setlength{\tabcolsep}{3pt}
\caption{R2 private-draw diagnostic. Values are means $\pm$ SE over 30 seeds.}
\label{tab:r2}
\begin{tabular}{@{}p{0.34\linewidth}rrrr@{}}
\toprule
Condition / reporting probe & gap & $\tau_d$ & detected & $W$ \\
\midrule
R1: realized schedule public & $+0.000 \pm 0.000$ & $12.0 \pm 0.0\;(=T)$ & $0\%$ & $400 \pm 0$ \\
R2: no schedule knowledge & $+0.050 \pm 0.000$ & $1.3 \pm 0.4$ & $100\%$ & $400 \pm 0$ \\
R2: independent schedule guess & $+0.036 \pm 0.002$ & $2.1 \pm 0.5$ & $100\%$ & $257 \pm 7$ \\
\bottomrule

\end{tabular}
\end{table}

\new{Both schedule-blind probes are detected by $T$ in every seed, with mean $\tau_d=\ResultRtwoBlindTau$ and $\ResultRtwoGuessTau$; Bonferroni-corrected times are identical in these cells. The limited conclusion is that OffAuditDrift's zero-gap guarantee depends on disclosure of the \emph{realized} schedule. This is neither an equilibrium solution nor an exhaustive characterization of behavior under uncertainty. Appendix~\ref{app:r2} documents the configuration and seed-level schedule records.}

\section{Limitations and discussion}
\label{sec:discussion}

\new{\paragraph{Scope.}This is a general continuous-compliance benchmark instantiated for AI-governance monitoring, not a mechanism unique to AI. The current scalar state omits model-version changes, interacting metrics, deployment-data shift, and compliance documentation. DSA and AI Act provisions motivate the setting and temporal analogies; Min-sample floor and Suspicion-escalation are proposed benchmark interventions, not asserted legal requirements.}

\new{\paragraph{Measurement and validation.}The simulator uses a scalar synthetic trajectory and sets the auditor's verified reference equal to the latent full-population quantity. Except in \texttt{HonestNoisyAuditee}, reports are generated directly rather than estimated from individual-level samples. Contextual parameterization from published DSA summary statistics is not platform-level validation, and the experiments provide no estimate of the prevalence of Delay, Attrition, definition shopping, or schedule exploitation in practice. The results therefore support design-level statements about specified probes and policies, not behavioral claims about regulated firms.}

\new{\paragraph{Incentives and dynamics.}There are no payoff functions, learning rules, best-response computations, remediation effects, or reputational consequences. A history-conditioned mapping can produce different actions as the trajectory evolves, but the agent does not revise that mapping during rollout. Consequently, the benchmark cannot establish that any supplied strategy is rationally optimal or that either side is in equilibrium.}

\new{\paragraph{Information and exposure accounts.}The headline schedule-aware results disclose the realized schedule. The R2 experiment changes that condition for one randomized policy and two simple probes; it is not a solution of the private-information problem. Auditor-side reference noise and richer secret or sequential policies remain unevaluated. Finally, $W$ and $C$ are stylized covered- and excluded-surface exposure accounts, not welfare estimates; domain transfer requires both substantive reparameterization and validation.}

\new{\paragraph{Extensions.}The interface can host vector-valued metrics, noisy auditor references, platform-derived trajectories, explicit costs and penalties, and Bayesian or sequential audit policies. These are directions for extending the evaluation surface, not results delivered here.}

\singlespacing
\printbibliography

\appendix
\doublespacing

\section{Implementation details}
\label{app:implementation}

\subsection{Reproduction commands}

To reproduce Table~\ref{tab:headline} on a fresh environment:

\begin{verbatim}
uv venv && uv pip install -e .
uv run python -m audit_gaming_benchmark.run \
    --config experiments/configs/default.yaml --out results/default
uv run python -m audit_gaming_benchmark.run \
    --config experiments/configs/attrition.yaml --out results/attrition
uv run python scripts/r2_offauditdrift.py \
    --config experiments/configs/r2-private-schedule.yaml \
    --out results/r2-offauditdrift/r2_offauditdrift.json
uv run pytest -q
\end{verbatim}

The headline runs each write a \verb|sweep.json| with per-cell aggregates. The R2 output additionally retains every environment seed, policy seed, realized audit schedule, and independently guessed schedule where applicable. New strategies and policies are added by implementing the respective \verb|Auditee| / \verb|Auditor| protocol and registering with the runner.

\subsection{R2 output schema}
\label{app:r2}

The R2 experiment uses the headline horizon and 30 environment seeds but keeps the Scheduled-random policy's realized draw private. The aggregate results appear in Table~\ref{tab:r2}. The output file \texttt{results/r2-offauditdrift/r2\_offauditdrift.json} records the configuration and aggregate rows, then stores every environment seed, policy seed, realized audit schedule, independently guessed schedule where applicable, and per-seed metric record. The regression suite pins all three aggregate rows and checks this seed metadata.

\subsection{Code structure}

The benchmark is a single Python package (\texttt{audit\_gaming\_benchmark}) with five source modules:

\begin{itemize}
  \item \texttt{env.py}: \texttt{AuditEnv} (the $T$-round audit environment), \texttt{Trajectory} (per-round outcomes container), and \texttt{Auditor} / \texttt{Auditee} structural \texttt{Protocol}s.
  \item \texttt{strategies.py}: \texttt{HonestAuditee}, \texttt{HonestNoisyAuditee}, \texttt{DelayStrategy}, \texttt{DriftStrategy}, \texttt{CherryPickStrategy}, \texttt{AttritionStrategy}, \texttt{OffAuditDriftStrategy}.
  \item \texttt{policies.py}: \texttt{OneShotPolicy}, \texttt{PeriodicPolicy}, \texttt{SurprisePolicy} (the implementation listing for the Scheduled-random policy of \S\ref{sec:benchmark}; the legacy class name is retained for backward compatibility with prior config files), \texttt{MinSampleFloorPolicy}, \texttt{SuspicionEscalationPolicy}.
  \item \texttt{metrics.py}: \texttt{gaming\_gap}, \texttt{detection\_threshold}, \texttt{time\_to\_detection}, \texttt{covered\_surface\_exposure}, \texttt{excluded\_surface\_exposure}, and \texttt{all\_metrics}. The older names \texttt{welfare\_loss} and \texttt{coverage\_loss} remain explicit compatibility aliases.
  \item \texttt{run.py}: YAML-driven sweep runner with multi-seed averaging.
\end{itemize}

\subsection{Auditor / Auditee protocol}

Both interfaces are deliberately minimal so that user-supplied strategies and policies plug in without modification:
\begin{itemize}
  \item \texttt{Auditee.choose\_report(t, true\_metric, audit\_history) -> (reported, n\_t)}: returns the reported metric and sample size for round $t$. Stateful strategies (e.g.\ \texttt{DelayStrategy}, which buffers past true metrics) maintain state on the instance.
  \item \texttt{Auditor.audit\_this\_round(t, T, history) -> bool}: returns whether to audit round $t$. Static policies ignore \texttt{history}; Suspicion-escalation uses it to implement the fixed history-conditioned mapping.
\end{itemize}

\subsection{Detection rule}

The implementation of equation~\eqref{eq:detection-rule}:
\begin{verbatim}
def detection_threshold(reported_p, n, epsilon=0.0, z=1.96):
    p = max(0.0, min(1.0, reported_p))
    se = math.sqrt(p * (1 - p) / max(n, 1))
    return epsilon + z * se
\end{verbatim}
At $p=0.5$ and $z=1.96$, this gives $\sim\ResultThresholdFull$ for $n=1000$, $\sim \ResultThresholdSmall$ for $n=100$. Detection fires when $|\hat m_t - m_t|$ exceeds this threshold on an audited round.

\subsection{Configuration schema}

Configurations are YAML; the runner reads \texttt{env}, \texttt{seeds}, \texttt{detection\_epsilon}, \texttt{strategies}, and \texttt{policies}. Each strategy/policy entry is a \texttt{\{name, params\}} pair where \texttt{name} keys into a class registry. New strategies and policies can be registered by adding to the dictionaries in \texttt{run.py}.

\subsection{Test coverage}

The 23-test pytest suite covers environment determinism, each strategy's defining behavior, policy cadence and history conditioning, metric completeness and legacy-alias agreement, multi-seed aggregation, and the widening of the Wald threshold as $n_t$ shrinks. Regression tests pin the Cherry-pick $\times$ \{Periodic, Min-sample floor\} cells, the corrected Drift and Attrition Suspicion-escalation cells, and all three R2 rows including their seed and schedule metadata; a smoke test validates the honest-noise output schema.

\section{Full experimental results}
\label{app:full-results}

\subsection{Full strategy-by-policy matrices}

Table~\ref{tab:full-default} (default config) and Table~\ref{tab:full-attrition} (attrition config) report every cell of the strategy-by-policy matrix on both configs. Cells are mean $\pm$ standard error over $S = 30$ seeds. Time-to-detection of $T = 12$ indicates ``never detected'' within the horizon.

\begin{table}[t]
\centering
\caption{Default config ($m_0=0.5$, $T=12$, 30 seeds). Auditor schedule seeds vary independently of auditee stochasticity. Excluded-surface exposure $C$ is suppressed for layout: non-attriting rows have $C=0$, while the Attrition block has mean $C\approx\ResultDefaultAttritionExcluded$. Table~\ref{tab:full-attrition} reports $C$ under the attrition config.}
\label{tab:full-default}
\small
\begin{tabular}{llrrr}
\toprule
\textbf{Strategy} & \textbf{Policy} & \textbf{Gap} & \textbf{$\tau_d$} & \textbf{$W$} \\
\midrule
Honest & One-shot & $\phantom{+}0.000 \pm 0.000$ & $12.0$ & $0$ \\
Honest & Periodic & $\phantom{+}0.000 \pm 0.000$ & $12.0$ & $0$ \\
Honest & Scheduled-random & $\phantom{+}0.000 \pm 0.000$ & $12.0$ & $0$ \\
Honest & Suspicion-escalation & $\phantom{+}0.000 \pm 0.000$ & $12.0$ & $0$ \\
Honest & Min-sample floor & $\phantom{+}0.000 \pm 0.000$ & $12.0$ & $0$ \\
Delay & One-shot & $-0.005 \pm 0.005$ & $10.1$ & $225 \pm 14$ \\
Delay & Periodic & $-0.002 \pm 0.002$ & $6.4$ & $163 \pm 12$ \\
Delay & Scheduled-random & $+0.004 \pm 0.002$ & $7.2$ & $169 \pm 10$ \\
Delay & Suspicion-escalation & $-0.003 \pm 0.003$ & $8.7$ & $164 \pm 12$ \\
Delay & Min-sample floor & $-0.002 \pm 0.002$ & $6.4$ & $163 \pm 12$ \\
Drift & One-shot & $+0.050 \pm 0.000$ & $5.0$ & $550 \pm 0$ \\
Drift & Periodic & $+0.050 \pm 0.000$ & $2.0$ & $400 \pm 0$ \\
Drift & Scheduled-random & $+0.050 \pm 0.000$ & $1.3$ & $400 \pm 0$ \\
Drift & Suspicion-escalation & $+0.050 \pm 0.000$ & $4.0$ & $200 \pm 0$ \\
Drift & Min-sample floor & $+0.050 \pm 0.000$ & $2.0$ & $400 \pm 0$ \\
Cherry-pick & One-shot & $+0.053 \pm 0.005$ & $5.7$ & $507 \pm 15$ \\
Cherry-pick & Periodic & $+0.045 \pm 0.003$ & $2.7$ & $375 \pm 13$ \\
Cherry-pick & Scheduled-random & $+0.045 \pm 0.002$ & $2.3$ & $379 \pm 14$ \\
Cherry-pick & Suspicion-escalation & $+0.044 \pm 0.002$ & $4.4$ & $236 \pm 19$ \\
Cherry-pick & Min-sample floor & $+0.045 \pm 0.003$ & $2.7$ & $375 \pm 13$ \\
Attrition & One-shot & $+0.002 \pm 0.002$ & $12.0$ & $1 \pm 1$ \\
Attrition & Periodic & $+0.001 \pm 0.001$ & $12.0$ & $1 \pm 1$ \\
Attrition & Scheduled-random & $+0.001 \pm 0.001$ & $12.0$ & $1 \pm 1$ \\
Attrition & Suspicion-escalation & $+0.001 \pm 0.001$ & $12.0$ & $0$ \\
Attrition & Min-sample floor & $+0.001 \pm 0.001$ & $11.7$ & $1 \pm 1$ \\
OffAuditDrift & One-shot & $\phantom{+}0.000 \pm 0.000$ & $12.0$ & $550 \pm 0$ \\
OffAuditDrift & Periodic & $\phantom{+}0.000 \pm 0.000$ & $12.0$ & $400 \pm 0$ \\
OffAuditDrift & Scheduled-random & $\phantom{+}0.000 \pm 0.000$ & $12.0$ & $400 \pm 0$ \\
OffAuditDrift & Suspicion-escalation & $\phantom{+}0.000 \pm 0.000$ & $12.0$ & $500 \pm 0$ \\
OffAuditDrift & Min-sample floor & $\phantom{+}0.000 \pm 0.000$ & $12.0$ & $400 \pm 0$ \\
\bottomrule

\end{tabular}
\end{table}

\begin{table}[t]
\centering
\caption{Attrition config ($m_0=0.30$, $T=12$, 30 seeds). Attrition activates every round; non-attriting strategies have $C=0$. Min-sample floor detects Attrition at $\tau_d=\ResultAttritionFloorTau$, while Suspicion-escalation does not detect it within the horizon ($\tau_d=\ResultAttritionSuspicionTau$), illustrating the two axes of Design Lemma~\ref{prop:no-static}. OffAuditDrift remains undetected in the disclosed-schedule headline condition.}
\label{tab:full-attrition}
\small
\begin{tabular}{llrrrr}
\toprule
\textbf{Strategy} & \textbf{Policy} & \textbf{Gap} & \textbf{$\tau_d$} & \textbf{$W$} & \textbf{$C$} \\
\midrule
Honest & One-shot & $\phantom{+}0.000 \pm 0.000$ & $12.0$ & $0$ & $0$ \\
Honest & Periodic & $\phantom{+}0.000 \pm 0.000$ & $12.0$ & $0$ & $0$ \\
Honest & Scheduled-random & $\phantom{+}0.000 \pm 0.000$ & $12.0$ & $0$ & $0$ \\
Honest & Suspicion-escalation & $\phantom{+}0.000 \pm 0.000$ & $12.0$ & $0$ & $0$ \\
Honest & Min-sample floor & $\phantom{+}0.000 \pm 0.000$ & $12.0$ & $0$ & $0$ \\
Delay & One-shot & $-0.005 \pm 0.005$ & $9.9$ & $225 \pm 14$ & $0$ \\
Delay & Periodic & $-0.002 \pm 0.002$ & $5.1$ & $163 \pm 12$ & $0$ \\
Delay & Scheduled-random & $+0.004 \pm 0.002$ & $6.7$ & $169 \pm 10$ & $0$ \\
Delay & Suspicion-escalation & $-0.003 \pm 0.003$ & $8.5$ & $164 \pm 12$ & $0$ \\
Delay & Min-sample floor & $-0.002 \pm 0.002$ & $5.1$ & $163 \pm 12$ & $0$ \\
Drift & One-shot & $+0.050 \pm 0.000$ & $5.0$ & $550 \pm 0$ & $0$ \\
Drift & Periodic & $+0.050 \pm 0.000$ & $2.0$ & $400 \pm 0$ & $0$ \\
Drift & Scheduled-random & $+0.050 \pm 0.000$ & $1.3$ & $400 \pm 0$ & $0$ \\
Drift & Suspicion-escalation & $+0.050 \pm 0.000$ & $4.0$ & $200 \pm 0$ & $0$ \\
Drift & Min-sample floor & $+0.050 \pm 0.000$ & $2.0$ & $400 \pm 0$ & $0$ \\
Cherry-pick & One-shot & $+0.053 \pm 0.005$ & $5.7$ & $507 \pm 15$ & $0$ \\
Cherry-pick & Periodic & $+0.045 \pm 0.003$ & $2.7$ & $375 \pm 13$ & $0$ \\
Cherry-pick & Scheduled-random & $+0.045 \pm 0.002$ & $2.3$ & $379 \pm 14$ & $0$ \\
Cherry-pick & Suspicion-escalation & $+0.044 \pm 0.002$ & $4.4$ & $236 \pm 19$ & $0$ \\
Cherry-pick & Min-sample floor & $+0.045 \pm 0.003$ & $2.7$ & $375 \pm 13$ & $0$ \\
Attrition & One-shot & $+0.050 \pm 0.000$ & $12.0$ & $54 \pm 0$ & $534 \pm 4$ \\
Attrition & Periodic & $+0.050 \pm 0.000$ & $12.0$ & $39 \pm 0$ & $534 \pm 4$ \\
Attrition & Scheduled-random & $+0.049 \pm 0.001$ & $12.0$ & $40 \pm 0$ & $534 \pm 4$ \\
Attrition & Suspicion-escalation & $+0.049 \pm 0.001$ & $12.0$ & $20 \pm 0$ & $534 \pm 4$ \\
Attrition & Min-sample floor & $+0.050 \pm 0.000$ & $\mathbf{2.0}$ & $39 \pm 0$ & $534 \pm 4$ \\
OffAuditDrift & One-shot & $\phantom{+}0.000 \pm 0.000$ & $12.0$ & $550 \pm 0$ & $0$ \\
OffAuditDrift & Periodic & $\phantom{+}0.000 \pm 0.000$ & $12.0$ & $400 \pm 0$ & $0$ \\
OffAuditDrift & Scheduled-random & $\phantom{+}0.000 \pm 0.000$ & $12.0$ & $400 \pm 0$ & $0$ \\
OffAuditDrift & Suspicion-escalation & $\phantom{+}0.000 \pm 0.000$ & $12.0$ & $500 \pm 0$ & $0$ \\
OffAuditDrift & Min-sample floor & $\phantom{+}0.000 \pm 0.000$ & $12.0$ & $400 \pm 0$ & $0$ \\
\bottomrule

\end{tabular}
\end{table}

\subsection{Sensitivity analyses}

\begin{figure}[t]
\centering
\includegraphics[width=\linewidth]{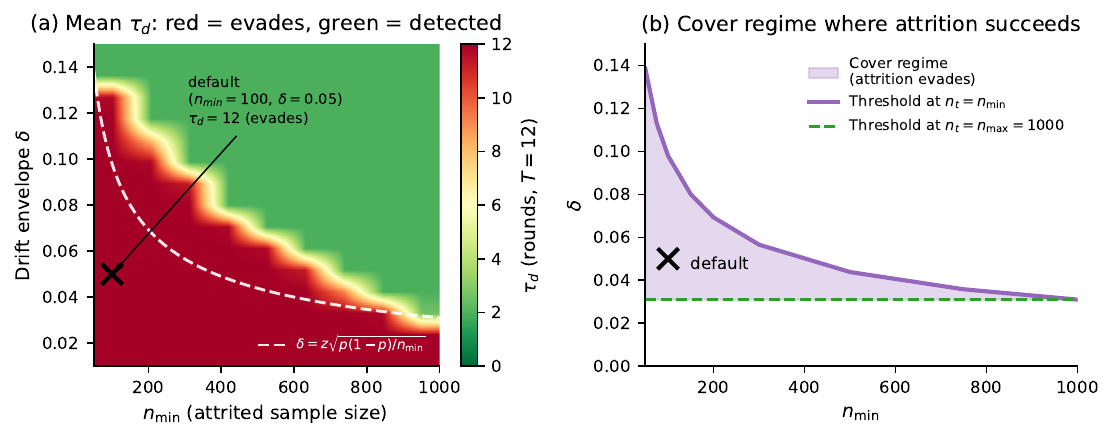}
\caption{\textbf{Cover regime in $(n_{\min}, \delta)$ space, periodic policy, $m_0 = 0.30$, $T = 12$, 30 seeds per cell.}
(a) Heat-map of mean time-to-detection $\tau_d$ under uncorrected Wald-CI. Red regions correspond to attrition-evasion ($\tau_d = T$), green regions to fast detection. The default operating point ($n_{\min} = 100$, $\delta = 0.05$) sits firmly in the cover regime, but the boundary is steep --- attrition is sensitive to both parameters.
(b) Cover regime as Design Lemma~\ref{prop:no-static} characterizes it: the shaded band is the set of $\delta$ values that exceed the full-sample threshold (green dashed line at $\ResultThresholdFull$) but fit within the small-sample threshold at each $n_{\min}$ (purple). The default sits inside the band; reducing the auditee's $n_{\min}$ flexibility (e.g., requiring reports on at least $500$ verified individuals) closes the regime for $\delta=0.05$.}
\label{fig:sensitivity}
\end{figure}

\paragraph{Cover-regime sensitivity in $(n_{\min}, \delta)$ space.} Figure~\ref{fig:sensitivity} maps the cover regime under the periodic policy.

\begin{figure}[t]
\centering
\includegraphics[width=\linewidth]{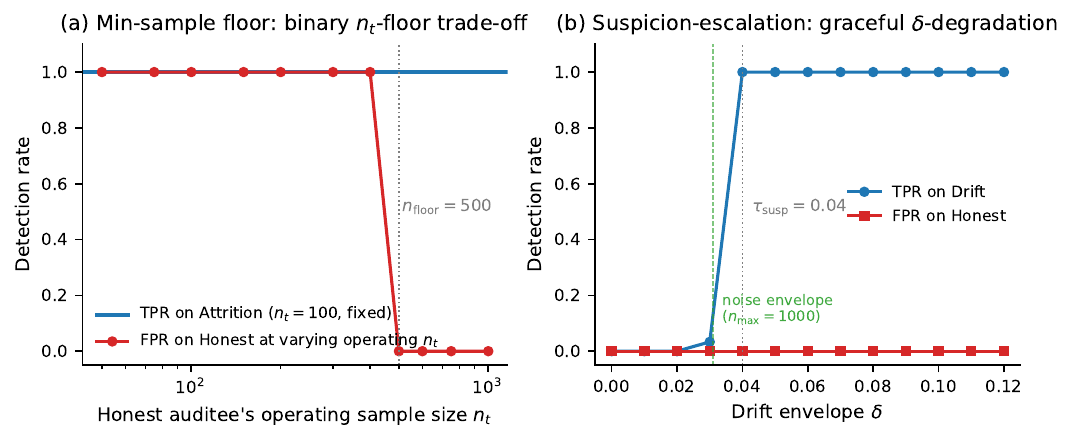}
\caption{\textbf{Sensitivity curves for the two intervention baselines, 30 seeds per point.}
(a) Min-sample floor: TPR on Attrition (fixed $n_t = 100$) is $1.0$ for any $n_{\text{floor}} > 100$ (the policy fires under strict inequality $n_t < n_{\text{floor}}$); FPR on a legitimate Honest auditee operating at sample size $n_t$ collapses step-wise at $n_t = n_{\text{floor}} = 500$. The trade-off appears structurally binary in this configuration: a floor that conditions only on the level of $n_t$ exposes the same signal under strategic attrition and under a legitimately small operating population, suggesting why richer history-conditioned frontiers using $n_t$ \emph{trajectories} (not just levels) are the natural follow-on direction.
(b) Suspicion-escalation: TPR on Drift rises sharply as $\delta$ crosses the noise envelope at $n_{\max}$ ($\approx \ResultThresholdFull$, green dashed), reaching $1.0$ above the suspicion threshold ($\tau_{\text{susp}} = 0.04$); FPR is zero in panel~(b) because the curve uses \texttt{HonestAuditee} ($\hat m_t = m_t$ exactly); under \texttt{HonestNoisyAuditee} per-round uncorrected FPR is $\approx 0.05$ and family-wise Bonferroni $\approx \ResultFprPeriodicBonf$ (Appendix~\ref{app:fpr}). The policy has a clean operating regime above the noise envelope and degrades gracefully below it.}
\label{fig:pareto}
\end{figure}

\paragraph{Intervention sensitivity.} Figure~\ref{fig:pareto} maps the operating regimes of the two intervention baselines.

The two panels expose orthogonal failure modes: Min-sample floor's trade-off is binary in $n_t$ (the TPR/FPR axes coincide on the same threshold variable); Suspicion-escalation's trade-off is graceful in $\delta$ (TPR and FPR are governed by independent variables). A richer history-conditioned frontier should combine both axes --- conditioning on $(n_t \text{ trajectory}, \text{gap history})$ jointly --- and is the natural follow-on (\S\ref{sec:discussion}).

\paragraph{Drift envelope.} The Drift strategy is parameterized by $\delta \in [0,0.5]$. As $\delta$ shrinks, the gap narrows linearly and time-to-detection grows once $\delta$ falls below the noise envelope at $n_{\max}$ (about $\ResultThresholdFull$). Below this threshold, non-detection is an ordinary low-signal case outside Design Lemma~\ref{prop:no-static}'s cover interval; Case~2 instead requires a drift above the full-sample threshold that is hidden only after $n_t$ shrinks. At $\delta \geq 0.05$, Scheduled-random detects on its first audited round under both configs.

\paragraph{Detection tolerance $\epsilon$.} Setting $\epsilon = 0.05$ widens the threshold by an additive $0.05$, allowing both Drift and Attrition strategies at $\delta = 0.05$ to escape all three policies. In the benchmark baseline, we set $\epsilon=0$; this is a modeling choice, not a description of current regulatory practice.

\paragraph{Attrition threshold $\tau$.} Setting $\tau=0.5$ in the default config causes Attrition to activate on roughly half the rounds. The resulting gap averages zero-gap and $+0.05$ rounds, producing intermediate covered-surface exposure; detection still fails because activated rounds reduce $n_t$ enough to remain inside the acceptance band.

\paragraph{Sample-size ratio.} The default contrast $n_{\min} / n_{\max} = 0.1$ widens the threshold from $\sim\ResultThresholdFull$ to $\sim\ResultThresholdSmall$, which is enough to swallow $\delta = 0.05$. More generally, the boundary follows the Wald condition $\delta \leq h(\hat m_t,n_{\min})$ and therefore varies with the reported trajectory, as Figure~\ref{fig:sensitivity} shows. At $n_{\min} = 500$, the threshold is only $\sim\ResultThresholdFloor$, and a $\delta = 0.05$ drift is still detected. The benchmark's sample-size contrast is deliberately chosen to place the stress test in a regime where attrition is a meaningful threat; users can choose weaker contrasts.

\subsection{Trajectory visualizations}

For each cell, the JSON output includes the per-round trajectory: $\{t, m_t, \hat m_t, n_t, \text{audited}_t\}$ for $t = 0, \dots, T-1$. Visualization scripts that produce time-series plots, gap-vs-detection-band overlays, and covered-surface exposure accumulation curves are provided in \verb|experiments/figures/| and produce camera-ready PDFs from the JSON.

\subsection{Compute requirements}

A full 30-seed sweep across the $5\times 5$ gaming-strategy-by-policy matrix on each config (50 gaming cells across both configs, plus 10 Honest baseline rows; $60$ cells total) completes in well under a minute on a 2024 MacBook (Python 3.11, NumPy 1.26, no GPU); a single-seed cell runs in under a second. The benchmark is intentionally lightweight; the analytical interest is in policy and strategy design, not in training neural surrogates. Future extensions with neural auditee policies will require GPU but are out of scope for the current submission.

\section{Bonferroni audit-count accounting and scheduled-random schedule-seed handling}
\label{app:bonferroni-K}

\paragraph{Fixed-cadence policies.} For fixed schedules $K = |\mathcal{A}_\pi|$, the committed audit-set cardinality: $K = 1$ for One-shot and $K = 4$ for Periodic, Scheduled-random, and Min-sample floor.

\paragraph{Suspicion-escalation.} For the history-conditioned policy we compute Bonferroni thresholds from each trajectory's \emph{realized} $K$. The base cadence is $\{4,8\}$, but a gap exceeding the raw suspicion threshold at $t=4$ causes audits at $\{4,5,\ldots,11\}$ and hence $K=\ResultSuspicionDriftK$. Drift and Attrition in the active-attrition configuration follow this path. Cherry-pick yields mean $K=\ResultSuspicionCherryK\pm\ResultSuspicionCherryKSe$, Delay $\ResultSuspicionDelayK\pm\ResultSuspicionDelayKSe$, and Attrition in the default configuration $\ResultSuspicionAttritionDefaultK\pm\ResultSuspicionAttritionDefaultKSe$ because activation varies across seeds. Honest and OffAuditDrift remain at $K=\ResultSuspicionHonestK$ and $K=\ResultSuspicionOffAuditK$, respectively. The raw-gap trigger is distinct from the Wald decision threshold: active Attrition escalates to $K=\ResultSuspicionAttritionActiveK$ yet remains undetected because its $0.05$ gap lies inside the wider band at $n_t=100$. A sequential or alpha-spending alternative would be tighter but is deferred to follow-on work.

\paragraph{Scheduled-random.} The Scheduled-random policy's audit schedule seed is varied independently of the auditee stochasticity seed (offset by $1009 \cdot s$) so each trial draws an independent committed schedule. Prior versions reused the YAML-configured \texttt{SurprisePolicy} seed across all trials, which collapsed the audit-schedule distribution to a single committed schedule and produced misleading point estimates; the offset removes that confound.

\section{Implementation considerations for precision floors}
\label{app:adoption-barriers}

\new{Min-sample floor is a proposed benchmark intervention, not a requirement attributed to the DSA or AI Act. The sensitivity analysis exposes three implementation considerations for any level-conditioned evidence floor; it does not establish why current regulators have or have not adopted one:}

\begin{itemize}
  \item \textbf{(i) Legitimately small populations.} A fixed floor also flags compliant reports from genuinely small operating populations, so the intervention requires a rule for distinguishing strategic exclusion from domain size.
  \item \textbf{(ii) Threshold calibration.} The chosen $n_{\text{floor}}$ changes both detectability and the set of reports declared insufficient; the benchmark value is illustrative rather than normatively justified.
  \item \textbf{(iii) Domain heterogeneity.} Appropriate evidence sizes depend on the monitored population, metric, and decision context, limiting direct transfer of a single numerical floor.
\end{itemize}

The \texttt{MinSampleFloor} baseline demonstrates a design change outside the static temporal-coverage class. Figure~\ref{fig:pareto}(a) shows the resulting level-based trade-off in $n_t$, motivating future policies that condition on evidence trajectories or richer context rather than a single threshold.

\section{Honest-noise false-positive calibration}
\label{app:fpr}

We calibrate the noise-aware Wald rule under honest report-side sampling noise. \texttt{HonestNoisyAuditee} reports $\hat m_t \sim \mathrm{Binomial}(n_t,m_t)/n_t$ at $n_t=n_{\max}=1000$, while the auditor observes $m_t$ exactly. Table~\ref{tab:fpr} reports per-round and family-wise false-positive rates across 200 seeds at $\alpha=0.05$. Auditor-side reference noise is not modeled.

\begin{table}[h]
\centering
\footnotesize
\caption{Empirical false-positive rates under \texttt{HonestNoisyAuditee} (200 seeds, $T=12$, $n_t=1000$). ``Per-round'' counts firings per audited round; ``FWER'' counts seeds with at least one firing. Rows are generated from \texttt{results/honest\_noisy\_fpr.json}.}
\label{tab:fpr}
\begin{tabular}{lrrrrr}
\toprule
Policy & $\bar K$ & per-round uncorr & per-round Bonf & FWER uncorr & FWER Bonf \\
\midrule
One-shot & $1.0$ & $0.050$ & $0.050$ & $0.050$ & $0.050$ \\
Periodic & $4.0$ & $0.044$ & $0.010$ & $0.165$ & $0.040$ \\
Scheduled-random & $4.0$ & $0.054$ & $0.016$ & $0.195$ & $0.065$ \\
Suspicion-escalation & $2.0$ & $0.049$ & $0.017$ & $0.095$ & $0.035$ \\
Min-sample floor & $4.0$ & $0.044$ & $0.010$ & $0.165$ & $0.040$ \\
\bottomrule

\end{tabular}
\end{table}

Per-round uncorrected rates span $\ResultFprPerRoundMin$--$\ResultFprPerRoundMax$, around nominal $\alpha=0.05$. For the four-audit policies, uncorrected family-wise rates span $\ResultFprFourAuditMin$--$\ResultFprFourAuditMax$, while Bonferroni-corrected rates do not exceed $\ResultFprBonfMax$. Min-sample floor matches Periodic because $n_t=1000>n_{\text{floor}}=500$ in the honest baseline, so the floor condition never fires.

\end{document}